\documentstyle[11pt]{article}

\begin{document}

\begin{normalsize}

\begin{center}

{\bf Kolmogorov Complexity, String Information, Panspermia and the Fermi
Paradox }

\vspace{0.2in}

\noindent V.G.Gurzadyan

\end{center}

\vspace{0.1in}

\noindent 
Dipartimento di Fisica, Universit\`a ``La Sapienza", Rome, 
and Yerevan Physics Institute

\vspace{0.2in}

{\bf Abstract} - 
Bit strings rather than byte files can be a mode of transmission both for intelligent
signals and for travels of extraterrestrial life.
Kolmogorov complexity, i.e. the minimal length of a binary coded string completely defining a system,
can then, due to its universality, become a key concept in the strategy of the search of 
extraterrestrials.  
Evaluating, for illustration, the Kolmogorov complexity of the human genome, one 
comes to an unexpected conclusion that a low complexity compressed string 
- analog of Noah's ark - will enable the recovery of the totality of terrestrial life.   
The recognition of bit strings of various complexity up to incompressible Martin-L\"{o}f 
random sequences, will require a 
different strategy for the analysis of the cosmic signals. 
The Fermi paradox "Where is Everybody?" can be viewed under in the light of such information panspermia, i.e.
a Universe full of traveling life streams.

\section{Introduction}

Diametrically opposing views on the existence of advanced civilizations are mainly due to the
uncertainty, to what extent features of our civilization can be attributed to
extraterrestrials. For example, why cannot a civilization be developed
on the basis of elementary particles in entangled quantum states 
instead of atoms or molecules? Various strategies for the search
of advanced civilizations range from detection of electromagnetic
signals \cite{CM} to the search of physical artefacts in our vicinity \cite{RW}.

My discussion is based on outlining the information aspect of the carriers of life.
The basic idea is to send a file containing all the information on a system,
up to such complex systems as human beings. The amount of information then, of course,
can be quite big. The size of the package will decrease drastically if, instead
of sending a file (byte sequences), one sends the program (bit strings) which is able to recover it.
Then, the Kolmogorov complexity, which is the minimal length of
the program defining a system \cite{K}, due to its universality
will act as the quantitative descriptor of the messages.

The recent progress in the 
deciphering of the human genome is used below to illustrate these ideas
(Thus far, sequences of 
three mammalian 
genomes are studied with reasonable accuracy: the complete human genome sequence was published 
in 2003 (the draft in 2001), 
the draft genomes of the mouse and rat were published in 2002 and 2004, respectively \cite{gen}).  
Evaluating the Kolmogorov complexity of the human genome,
I arrive at an unexpected conclusion that low complexity strings will enable 
the complete recovering of the terrestrial life. 
If true, the methods of analysis of bit strings in the radiation arrived from space have to become an important goal.

\section{Kolmogorov complexity of the human genome}

Kolmogorov complexity, $K$, is defined as the minimal length of a
binary coded program (in bits) which is required to describe the
system $x$, i.e. will enable recovery of the initial system $x$ completely \cite{K}:
\begin{equation}
K(\phi(p),x) =min_{p:\phi=x} l(p),
\end{equation}
where $\phi(p,x)$ is a recursive i.e. algorithmically calculable
function, $l(p)$ is the length of the program $p$. 

The universality of the Kolmogorov complexity is due to the proof
by Kolmogorov that the complexities
defined by various Turing machines \cite{K} differ by no more than an additive constant $C$
$$
|K(\phi(p),x) - K(x\mid y)|\leq C,
$$
where the conditional complexity $K(x\mid y)$ of object $x$ when
the complexity of another object $y$ is known, is
\begin{equation}
K(x\mid y)=min\,\ l(p).
\end{equation}
The amount of information of object $x$ with respect to an object $y$ 
is evaluated from the complexities
$$
I(y:x)=K(x)- K(x\mid y).
$$
Complexities obtained by different algorithms differ from the asymptotic (minimal) one 
by another additive constant.
In other words, a system can be recovered from a compressed 
string almost independently on the computer and the algorithm.

Obviously, a repeat or periodic string has a low complexity, and it can be compressed more 
compactly than the chaotic one with random sequences. The precise complexity is usually unreachable 
for physical complex systems; however a value not too different from it can be estimated, as
for example, for the maps of Cosmic Microwave Background radiation \cite{G}. 

Let us estimate the complexity of the human genome. 
The human genome \cite{gen} contains 2.9 $10^9$ base pairs, those of the rat and mouse contain 
2.75 $10^9$ and 2.6 $10^9$ pairs, respectively. The number of predicted genes is about 23.000 
for the human and over 22.000 for the mouse genome. About 99\% of mouse genes are similar to 
the human ones, and of these 96\% have the same location for the mouse and human genome, 80\% of mouse 
gene (orthologues) are also the best match for human gene. 
Therefore, the complexity of the human genome has to be
\begin{equation}
K<10^{11}, 
\label{k}
\end{equation}
with correspondingly smaller value for the code carrier part.
Only less than 1\% of mouse proteins are specific to mice, 99\% per cent are shared with other mammals, and 98\% are 
shared with the humans, while 27\% are common to all animals and 23\% to all species, including bacteria.
Another feature of the mammalian genomes is the existence of repeat sequences.
Namely, 46\% of the human genome and 38\% of the mouse genome consists of recognizable interspersed repeats while only 2\% are the coding genes. The complexity of a string of length N is limited by $K(x)<N$,
while the fraction of sequences with 
$$K<N-m$$ 
is small if $m$ is sufficiently large, as it is for the human genome. 
The chromosomes of the three studied organisms, 23, 21 and 20 pairs for human, rats and mice, respectively, are related to each other by 280 large regions of sequence similarity. So, the conditional complexity of terrestial species is small once the complexity of human genome is known. (The similarities in the genomes are not only the quantitative indications for the common ancestors but also for the time periods of the divergence from the common path of evolution.) 

The energy, E, required to communicate $B$ bits of information is $E=BkT\, \ln2$, where $T$ is the temperature of the noise per bandwidth. A lower bound for the energy to transmit $B$ bits by an antenna can be evaluated by the formula \cite{RW}
\begin{equation}
E=8\, \ln2\, BS (\frac{D}{A})^2\simeq 10^6 (B/10^{11})(d/1pc)^2(R/150m)^{-4}\, erg.
\label{E}
\end{equation}
where $D$ is the communication distance, $d$, in units of the antenna's aperture, $A$ is the antenna's aperture, $R$, in units of the transmission wavelength, and $S$ is the noise spectral energy density. The Arecibo    
aperture, $R=150m$, and 3K antenna temperature is used for the normalization, so it is seen that larger antennae will enable the coverage of the Galaxy and even other galaxies within energy limits reasonable for our civilization in foreseen future.

Thus, the complexity of genomes of terrestrial organisms due to repeat sequences and common fractions is comparable to the human one and the resulting package can be transmitted to galactic distances.

\section{Network of von Neumann automata}

The self-reproduction of information carriers is an efficient
strategy for spreading over the Universe.  A simplified example of such
strategy can include sending N self-reproducing von Neumann cellular automata \cite{vN}, 
as suggested by Tipler \cite{T}. 
The automata would create duplicates of their own from the environment upon arrival 
at the destination, and send them in other N directions.  At a speed $0.001$c, an 
automaton will arrive to the nearest star situated 1pc away in $\tau=3,000$ years, and 
the time of creation of automata network within the Galaxy will be 
\begin{equation}
10^4 \tau = 30\, 000\, 000\, years
\end{equation}
At a speed $0.01c$, the Universe within the radius $10^{26}$ cm would be reached in the Hubble time. 

Once the network of von Neumann automata is created within Galaxy in such cosmologically short time scale, the transmission of information packages i.e. packed travellers, can be as commonplace events as air travels  on the Earth today. 
I do not discuss many traditionally debated issues, such as whether the civilizations would have other alternatives to the expansion in space, etc.

\section{Conclusion}

I advance the idea of propagation of the life not via files containing the information on them, but the
programs, i.e. coded strings defined by Kolmogorov complexity.  
Considering the Kolmogorov complexity of the human genome, I have shown that low-conditional-complexity 
strings are enough for the complete recovery of terrestrial life.  
The complexity of the coded strings, the existence of random sequences in the sense of Martin-L\"{o}f 
closely related to 
Hausdorff dimensionality, will require new methods in the studies of the cosmic signals,
and can eventually approach the solution to the Fermi paradox  "`Where is Everybody?" \cite{W}.

\end{normalsize}

\end{document}